\documentclass[aps,prl,preprint,groupedaddress,amsmath,showpacs]{revtex4-1}
\usepackage{hyperref}
\usepackage{slashed}
\usepackage{dsfont}
\def\ben{\begin{equation}}
\def\een{\end{equation}}
\def\bena{\begin{eqnarray}}
\def\eena{\end{eqnarray}}
\def\non{\nonumber}
\def\C{{\cal C}}
\def\psib{{\overline \psi}}
\def\bpsi{{\overline \psi}}
\newcommand{\N}{\tfrac{1}{N}}
\newcommand{\sN}{\tfrac{1}{N^2}}
\newcommand{\vac}{{\rm vac}}
\newcommand{\half}{\tfrac{1}{2}}
\newcommand{\e}{{\rm e}}

\begin{document}


\title{A small cosmological constant\\ due to non-perturbative quantum effects}


\author{Jan Holland}
\email[Email: ]{HollandJW1@cf.ac.uk}
\affiliation{School of Mathematics, Cardiff University\\ Senghennydd Rd., CF24 4AG, Cardiff, United Kingdom}
\author{Stefan Hollands}
\email[Email: ]{Hollandss@cf.ac.uk}
\affiliation{Institut f\"ur Theoretische Physik, Universit\"at Leipzig\\ Br\"uderstr. 16, Leipzig, D-04103, Germany}


\date{\today}

\begin{abstract}
We propose that the expectation value of the stress energy tensor of the Standard Model
should be given by $\langle T_{\mu \nu} \rangle = \rho_\vac\, \eta_{\mu\nu}$, with a vacuum
energy $\rho_\vac$ that differs from the usual ``dimensional analysis'' result by an exponentially small factor associated with non-perturbative effects. We substantiate our proposal by a rigorous analysis of a toy model, namely the 2-dimensional Gross-Neveu model. In particular, we address, within this model, the key question of the renormalization ambiguities affecting the calculation. The stress energy operator is constructed concretely via the operator-product-expansion. The non-perturbative factor in the vacuum energy is seen as a consequence of the facts that a) the OPE-coefficients have an analytic dependence on $g$, b) the vacuum correlations have a non-analytic (=non-perturbative) dependence on $g$, which we propose to be a generic feature of QFT.
Extrapolating our result from the Gross-Neveu model to the Standard Model, one would expect to find $\rho_\vac \sim \Lambda^4 \e^{-O(1)/g^2}$, where $\Lambda$ is an energy scale such as
$\Lambda = M_{\rm H}$, and $g$ is a gauge coupling such as $g^2/4\pi = \alpha_{\rm EW}$. The exponentially small factor due to non-perturbative effects could explain the ``unnatural'' smallness of this quantity.

\end{abstract}


\maketitle

\section{Introduction}

One of the major puzzles in modern cosmology is the origin of Dark Energy, and its apparently `unnatural' magnitude. Many, and very diverse, explanations have been proposed in this direction, see e.g.~\cite{carroll} for a review. Many of these proposals involve highly speculative features such as hypothetical new fields or dynamical mechanisms that have neither been observed, nor have been explored thoroughly from the theoretical viewpoint.

A very economical, and perhaps the most natural, hypothesis is that Dark Energy is simply quantum field theoretic vacuum energy. In other words, it is simply the expectation value of the quantum field theoretic stress energy operator, $\langle T_{\mu\nu} \rangle$, of the Standard Model of particle physics. The quantum state should in principle contain the approximately $10^{80}$ hadronic particles in the universe distributed onto stars, galaxies, dust clouds, etc. But for the problem at hand, we are not really interested in the detailed functional form of $\langle T_{\mu\nu} \rangle$ on smaller scales arising from these features, but rather in the contribution from the vacuum itself, in particular since the universe is mostly empty. Hence, one may take the state to be the vacuum state. Also, although our universe is expanding, its expansion rate is so small compared to the scales occurring in particle physics that we may safely do our analysis in Minkowski spacetime. Since the Minkowski vacuum state is Poincar\' e invariant, the vacuum expectation value must automatically have the form $\langle T_{\mu\nu} \rangle = \rho_\vac \eta_{\mu\nu}$ of a cosmological constant.

A natural guess for $\rho_\vac$ in the Standard Model, based essentially on dimensional analysis, is $\rho_\vac \sim \Lambda^4$, where
$\Lambda$ is a characteristic energy scale of the Standard Model, such as,
perhaps, $\Lambda = M_{\rm H} \sim 125\, {\rm GeV}$. This is well-known to be in striking
conflict\footnote{The same conclusion is drawn if $M_{\rm H}$ is replaced by other
natural scales associated with the Standard Model such as the scale of electro-weak
symmetry breaking.} with the observed value~\cite{particledata, wmap, amendola} of $\rho_\vac \sim (10^{-12}\, {\rm GeV})^4$. In this paper, we propose that a proper QFT-calculation of $\rho_\vac$ should rather result in a value of the type $\rho_\vac \sim \Lambda^4 \e^{-O(1)/g^2}$, with $\Lambda$ a typical energy scale of the Standard Model, such as perhaps $\Lambda = M_{\rm H}$, and with $g$ a gauge coupling such as perhaps $g^2/4\pi = \alpha_{\rm EW} \sim \frac{1}{137}$. This can give the right order of magnitude for $\rho_\vac$ for a suitable constant $O(1)$ of order unity, to be calculated in principle from the Standard Model. The essential point is that our proposal differs from dimensional analysis by an exponentially small, dimensionless factor, which we attribute to non-perturbative effects.

To justify our proposal rigorously, one would have to overcome the following two fundamental problems:

1) The huge complexity of the Standard Model, and in particular, the difficulty of making non-perturbative calculations.

2) The fact that, as is well-known, `the' stress energy operator, like any other `composite operator' in QFT, i.e. polynomial in the `basic fields', is an intrinsically ambiguous object.

Point 1) requires no further comment, except maybe that one cannot expect to be able to calculate a `form factor' like $\langle T_{\mu\nu} \rangle$ by perturbative methods. To substantiate our proposal in a clean setup, we therefore consider a toy model which is tractable, and at the same time displays some of the non-perturbative effects characteristic for the Standard Model. This model is the well-known Gross-Neveu model in two dimensions \cite{Gross:1974jv, abdalla} (we expect very similar results to hold also for the two dimensional $O(N)$ sigma model, treated along the lines of~\cite{novikov}). However, before we describe the Gross-Neveu model, let us say more clearly what we mean by 2). Given some quantum field operator $A$, we are free in general to make a field redefinition
\ben\label{amb}
A(x) \to \sum_B Z_A^B \cdot B(x)
\een
and to consider the right side as our new, equally legitimate (!), definition of that operator. In the context of standard renormalized perturbation theory around a Gaussian fixed point, the ambiguity can be attributed to the necessity of imposing ``renormalization conditions'', a change of which can be seen to correspond to field redefinitions~\eqref{amb}. The `mixing matrix' of complex numbers, $Z_A^B$, is somewhat restricted by various obvious requirements. For example, we want $\sum Z_A^B \cdot B$ to have the same tensor/spinor character as $A$. Also the field redefinition should not be in conflict with Poincar\'e invariance, and it should respect the quantum numbers of fields associated with any other symmetry of the theory. If we are near a Gaussian fixed point (e.g. in perturbation theory), we can naturally assign a dimension $\Delta_A$ to each composite operator, and the field redefinition should not increase the dimension, so $\Delta_B \le \Delta_A$ in the sum~\eqref{amb}. If the theory depends on a coupling constant $g$ (so that we not only have one QFT, but a 1-parameter family), then $Z_A^B(g)$ can be a function of $g$, but it is reasonable to require it should have a smooth dependence on $g$. Also, if $A(x)$ satisfies a differential relation such as a conservation law, then so should the right side of~\eqref{amb}. If $A=A^\dagger$ then $Z^A_{B^\dagger}=(Z_B^A)^*$, etc.

To illustrate these restrictions, suppose $A$ is a conserved current $J^\mu$ associated with a symmetry of the theory. If there is no other conserved current in the theory, then the only possible field redefinition is $J^\mu \to Z J^\mu$ for $Z$ real. The corresponding conserved charge $Q=\int J^0 d^3 x$ should furthermore generate the symmetry, $[Q,B(x)] = iq_B B(x)$, where $q_B$ is the charge quantum number of the operator $B$. Since $q_B$ is fixed, we must have $Z=1$ in this example. Thus, the current $J^\mu$ is uniquely defined as an operator. Consider
next the case when $A$ is the stress energy operator $T_{\mu\nu}$ of the theory. This operator should satisfy $\partial_\mu T^{\mu\nu} = 0$, so the stress tensor can only mix with other conserved operators that are symmetric tensors. A possible field redefinition is now
\ben\label{tredef}
T_{\mu\nu} \to Z \, T_{\mu\nu} + c \, \eta_{\mu\nu} {\bf 1} \ ,
\een
where ${\bf 1}$ is the identity operator, and $c$ a dimensionful constant. For example,
if the microscopic Lagrangian of the theory contains a single mass parameter, $M$, then $c \propto M^4$. Similarly to the previous example, we want $P_\mu = \int T^0{}_\mu d^3 x$ to generate translations, $[P_\mu, B(x)] = i\partial_\mu B(x)$, so
we must have $Z=1$. But, unfortunately, no restriction is obtained on the real constant $c$ that way. Since $\langle {\bf 1} \rangle = 1$, our field redefinition changes
$\langle T_{\mu\nu} \rangle \to \langle T_{\mu\nu} \rangle + c \eta_{\mu\nu}$, so we can set $\rho_\vac$ to any value we want. Furthermore, in a theory
depending on a coupling constant, $g$, we may let $c(g)$ be any (smooth) function of $g$ that we want, so we can even give the expected stress tensor an essentially arbitrary
dependence on the coupling constant. Therefore, unless we impose other reasonable conditions to cut down the ambiguity, we simply cannot predict what $\langle T_{\mu\nu} \rangle$ is within the framework of quantum field theory.

In order to motivate such a condition, we must better understand the true nature of `products' of operators in quantum field theory. The {\em only} natural definition
of product is in fact provided by the operator product expansion (OPE), which states
\ben\label{ope}
\left\langle A(x) B(0) \prod_i \phi(z_i) \right\rangle
\sim \sum_C \C_{AB}^C(x) \ \left\langle C(0) \prod_i \phi(z_i) \right\rangle \ .
\een
The $\phi(z_i)$ are ``spectator fields'', and the sum over the composite fields $C$ is organized by their dimension, in the sense that the numerical coefficients $\C_{AB}^C(x)$
are most singular in $x$ for the operator $C$ with the smallest dimension, and become more and
more regular as the dimension of $C$ increases. The $\sim$ sign means that if we subtract the partial sum up to a large dimension of $C$ from the right
side, then we get a quantity that goes to 0 fast as $x \to 0$. In this sense, the OPE is a short distance expansion.
The OPE coefficients encode the dynamics of the theory, and depend in particular on the coupling constants in the Lagrangian. We may indicate this by writing
$\C_{AB}^C(x;g)$, where $g$ is the (or possibly several) coupling constant. Clearly, if we make a $g$-dependent field redefinition~\eqref{amb} with
mixing matrix $Z_A^B(g)$, then the OPE coefficients will change accordingly.
Suppose, now, that there exists a definition of the composite fields such that $\C_{AB}^C(x;g)$ is an {\em analytic} function of $g$, i.e. has a {\em convergent}
Taylor expansion in $g$ for small, but finite, $g$. Then it is natural to allow only field redefinitions $Z_A^B(g)$ preserving this property, i.e. ones which are likewise {\em analytic}
in $g$. Therefore, for example, we would only be allowed to make a redefinition~\eqref{tredef} for an {\em analytic} function $c(g)$. Such analytic
field redefinitions could therefore not cancel out any non-analytic ($=$ non-perturbative) dependence on $g$ of the vacuum expectation value (VEV) $\langle T_{\mu\nu}\rangle$. Thus, if the theory has non-perturbative effects showing up in the VEV $\langle T_{\mu\nu}\rangle$, these cannot be removed by a, necessarily analytic, field redefinition. This leaves us, in principle, with the possibility of having an unambiguous, non-perturbatively small vacuum energy~\cite{hollands-wald}. The purpose of this paper is to substantiate this idea in the Gross-Neveu model in $d=2$.\\

\section{The model}

The massless (classically) Gross-Neveu model in $d=2$ dimensions is described by the Lagrangian \footnote{
Our conventions are (cf. \cite{abdalla}): signature $(+,-)$, Dirac matrices $\gamma^0 = \left(\begin{smallmatrix}
0 & -i \\
i & 0
\end{smallmatrix}\right)$, $\gamma^1 = \left(\begin{smallmatrix}
0 & i \\
i & 0
\end{smallmatrix}\right)
$,
Dirac conjugate $\bpsi = \psi^\dagger \gamma^0$
.}
\ben\label{Lagrange}
\mathcal{L}= N \left[ i \ \psib \slashed{\partial} \psi + \frac{g^2}{2} \ (\psib \psi )^{2} \right]\ ,
\een
where $\bpsi$ and $\psi$
are row/column vectors of $N$ flavors of a 2-component spinor field.
Relative to the usual presentation of the Lagrangian \cite{Gross:1974jv, abdalla}, the fields have been
rescaled by $1/\sqrt{N}$, which is convenient in view of the large $N$ limit taken later.
In the Lagrangian, and in similar expressions below, the flavor index is summed over in the obvious way. The expression for the classical stress-energy tensor per flavor, i.e. divided by $N$, is:
\ben
\label{tab}
T_{\mu\nu} =  \frac{i}{2} \ \psib \gamma_{\mu} \partial_{\nu} \psi + \frac{i}{2} \ \psib \gamma_{\nu} \partial_{\mu} \psi
 - \eta_{\mu \nu} \ \left[ i \ \psib \slashed{\partial} \psi + \frac{g^2}{2} \ (\psib \psi )^{2} \right] \ .
\een
The `t~Hooft coupling constant $g$ is dimensionless, and the model is conformally invariant at the classical level. By contrast, the corresponding quantum field theory is not conformally invariant, but exhibits the phenomenon of ``dynamical mass generation''. This means concretely that, at large space-like separation $x$, the 2-point correlation function has an exponential fall-off,
$
\langle  \bpsi(x) \psi(0) \rangle \sim
\exp(-K(g) \sqrt{-x^2/\ell^2})
$,
where $K(g)>0$ is a numerical constant, and where $\ell$ is a constant of dimension [length]. The dynamically generated mass
is accordingly given by $m(g) := K(g)/\ell$. The constant $\ell$ corresponds to a choice of ``units'' (fm, mm, km, etc.), which are obviously not provided
by the classical, scale invariant, Lagrangian~\eqref{Lagrange}. Its value is therefore best viewed as part of the definition of the quantum field theory.

The exponential fall-off
was rigorously proven, for sufficiently large but {\em finite} $N$, by \cite{Kopper:1993mj}.
The effect of mass generation is non-perturbative, in the sense that all $g$-derivatives of $K(g)$, corresponding to the various perturbation orders, vanish.
It was first discovered in a large $N$ analysis of the model by Gross and Neveu~\cite{Gross:1974jv}. The model becomes essentially solvable in this limit, and we have, in fact [compare eq.~\eqref{2pt}]
$
K(g) = \e^{-\pi/g^2} + O(\N)
$.
In spite of this characteristic non-perturbative dependence on $g$,
we will see that the OPE coefficients have a perfectly analytic dependence on $g$, see eqs.~\eqref{ope1},~\eqref{ope2},~\eqref{ope3}, as proposed in the previous section.
The full quantum field theory is defined by the collection of all $n$-point correlation functions of the {\em basic} fields $\psi, \overline \psi$, but we will only need the 2-~and 4-point functions. The 2-point function is (here and below we assume $x$ to be space-like):
\ben\label{2pt}
\langle  \psib_{\alpha} (x) \psi_{\beta} (0) \rangle = -\frac{(i\slashed{\partial}_{x} +m )_{\beta\alpha} }{2\pi}K_{0}(\sqrt{-x^2 m^{2}})+O(\N)
\een
where  $\alpha,\beta$ are spinor indices.
The 4-point function is written most conveniently as
\begin{widetext}
\bena\label{4pt}
&&\langle \bpsi_{\alpha}(x) \psi_\beta(0) \bpsi_\gamma(z_{1}) \psi_\delta(z_{2}) \rangle = \frac{-1}{2N}\int\frac{\text{d}^{2} p\,\text{d}^{2} q }{(2\pi)^{4}} \frac{[(\slashed{p}+m)(\slashed{q}+m)]_{\gamma\delta} \ \e^{i(z_{1}-x)p+i(x-z_{2})q} }{(p^{2}-m^{2})(q^{2}-m^{2})\, B(q-p)} \\
&&\times [(\slashed{q}-\slashed{p}+m-i\slashed{\partial}_{x})(m-i\slashed{\partial}_{x})]_{\alpha\beta}  \int_{0}^{1}\!\!\text{d}\alpha\  \frac{\sqrt{-x^{2}}K_{1}[\sqrt{-x^{2}(m^{2}-\alpha(1-\alpha)(q-p)^{2})}]\e^{ix(q-p)(\alpha-1)}}{m^{2}-\alpha(1-\alpha)(q-p)^{2}} \non \\
&&+\langle \bpsi_{\alpha}(x) \psi_\beta(0)\rangle\,  \langle\bpsi_\gamma(z_{1}) \psi_\delta(z_{2}) \rangle  -\frac{1}{N} \langle \bpsi_{\alpha}(x) \psi_\delta(z_{2})\rangle\,  \langle\bpsi_\gamma(z_{1}) \psi_\beta(0) \rangle + O(\sN)\, , \non
\eena
for our purposes. $K_{\alpha}$ are modified Bessel functions, and we use the short-hand
\ben
B(k):= \sqrt{\frac{4m^{2}-k^{2}}{-k^{2}}} \ln\frac{\sqrt{4m^{2}-k^{2}}+\sqrt{-k^{2}}}{\sqrt{4m^{2}-k^{2}}-\sqrt{-k^{2}}}\, .
\een
Note that the correlation functions have a {\em non-analytic} dependence on $g$ through
$m=\e^{-\pi/g}/\ell$.
Correlation functions of {\em composite} operators can be obtained from the correlation functions of the {\em basic fields} by means of the OPE. We will need the following OPE's in this paper:
\ben\label{ope1}
\bpsi(x) \psi(0) =
 O(\N)
 {\bf 1} +\bigg[
1- \frac{g^2}{2\pi}\log\left( \frac{-x^{2} \e^{2\Gamma_{\rm E}} }{4\ell^{2}}\right)
+ O(\N)
\bigg] \bpsi \psi(0) + \dots
\een
\ben
\label{ope2}
\bpsi(x) \slashed{\partial} \psi(0) =O(\N) {\bf 1}+O(\N) \,\bpsi\psi(0)+ \frac{g^2}{i} \bigg[
1- \frac{g^2}{2\pi}\log\left( \frac{-x^{2}\e^{2\Gamma_{\rm E}} }{4\ell^{2}}\right)
+ O(\N)
\bigg] (\bpsi \psi)^2(0) + \dots
\een
\bena
\label{ope3}
 \bpsi(x) \gamma_{(\mu} \partial_{\nu)} \psi(0) &=&\left[\frac{ -2x_{(\mu}x_{\nu)}}{i\pi\, x^{4} }+O(\N) \right] \mathbf{1}+O(\N) \,\bpsi\psi(0)+\!\!\bigg[1 + O(\N) \bigg] \bpsi \gamma_{(\mu} \partial_{\nu)} \psi(0)\non\\
  &-&
\bigg[ \frac{g^4 x_{(\mu}x_{\nu)}}{2\pi i \ x^{2}}  + O(\N)
\bigg] (\bpsi \psi)^2(0) + \dots
\eena
\end{widetext}
Here $\Gamma_{\rm E}$ is the Euler-Mascheroni constant, $t_{(\mu\nu)} = \half (t_{\mu\nu}+t_{\nu\mu}) - \half \eta_{\mu\nu} t_\sigma{}^\sigma$ is the symmetric traceless part and dots are terms of order $O(x)$. The OPE coefficients in these expressions were calculated using standard $\N$-expansion
techniques and e.g. the methods of~\cite{collins}. Terms not written 
explicitly are not needed later. We also have
$g^2(\bpsi \psi)^2 = i\,\bpsi \slashed{\partial} \psi$ as an operator equation --- in fact, one may consistently view this as the {\em definition} of the operator $(\bpsi \psi)^2$ --- which
is (formally) a consequence of the equation of motion. We see explicitly that the OPE coefficients are {\em analytic} in $g$, in contrast to the correlation functions.

\section{VEV of $T_{\mu\nu}$}

We would like to calculate the VEV of $T_{\mu\nu}$ [cf.~\eqref{tab}], which is evidently
a {\em composite operator}. VEV's of composite operators are calculated from the correlation
functions of the basic field $\bpsi, \psi$ by means of the OPE, and are subject to the
intrinsic renormalization ambiguities mentioned above. As a warm-up, let us illustrate the procedure for the VEV $\langle \bpsi \psi(0) \rangle$. First, we take an expectation value of eq.~\eqref{ope1}, solve for $\langle \bpsi \psi(0) \rangle$, and take $x \to 0$:
\ben
\langle \bpsi \psi(0) \rangle = \lim_{x \to 0} \frac{\langle \bpsi(x) \psi(0) \rangle -
 O(\N) \langle {\bf 1} \rangle}{1- \frac{g^2}{2\pi}\log\left( \frac{-x^{2} \e^{2\Gamma_{\rm E}} }{4\ell^{2}}\right)
+ O(\N)} \ .
\een
We now substitute eq.~\eqref{2pt} for the 2-point function,
and ignore terms of $O(\N)$. Making use of the standard expansion of the Bessel-function $K_0$
for small argument, we find
\ben
\langle \bpsi \psi(0) \rangle = \frac{-1}{g^2\ell} \e^{-\pi/g^2} + O(\N) \ .
\een
Of course, the VEV is the same at any other spacetime point $x$ by translation invariance.
Thus, we see that the VEV has a {\em non-analytic} dependence on $g$, and we cannot make the
VEV zero for all $g$ by any, necessarily {\em analytic}, field redefinition~\eqref{amb} of $\bpsi \psi$. In fact, the VEV shows that not only conformal-, but also (discrete) chiral symmetry is broken in the quantum theory.

Let us now determine the VEV of the stress tensor~\eqref{tab} by this method. We have to
be more careful here, because we need to make sure our definition of this composite operator obeys $\partial^\mu T_{\mu\nu} = 0$ as an operator equation. Our strategy is to define separately the composite operators appearing in formula~\eqref{tab} by the same method as just described. Their sum defines a composite operator, which actually turns out {\em not} to be conserved.
But fortunately we can add another operator of the same dimension (field redefinition) to it such that it now is conserved [up to order $O(\sN)$]. The resulting conserved operator is then the physical stress energy operator, which is seen to have a non-zero VEV. Let us now describe this in some more detail. Since we can consistently assume that $i\,\bpsi \slashed{\partial} \psi = g(\bpsi \psi)^2$ as an operator equation, we may simply write $T_{\mu\nu} = i\bpsi \gamma_{(\mu} \partial_{\nu)} \psi$. This composite operator is defined using the OPE~\eqref{ope3}. Since we would like to
check whether it is conserved {\em as an operator}, we need to calculate the divergence
$\langle \partial^\mu T_{\mu\nu}(0) \prod \bpsi(y_i) \prod \psi(z_j) \rangle$ inside a correlation function. Actually, it suffices to consider two spectator fields $\bpsi(z_1) \psi(z_2)$ inside the correlator. Solving eq.~\eqref{ope3} for $T_{\mu\nu}(0) = i\bpsi \gamma_{(\mu} \partial_{\nu)} \psi(0)$, inserting the result into the correlator in question, taking $x \to 0$, and using the well-known expansion for small argument of $K_1$ in~\eqref{4pt}, we get an expression for
$
\langle T_{\mu\nu}(0)  \bpsi(z_1) \psi(z_2) \rangle
$
in terms of the 4-point function~\eqref{4pt}.
We also need $\langle (\bpsi \psi)^2(0) \ \bpsi(z_1) \psi(z_2) \rangle$ which is obtained
in terms of the 4-point function~\eqref{4pt} in a similar way, using~\eqref{ope2} this time.  Then using the concrete form of~\eqref{4pt}, one derives, after a somewhat lengthy calculation, the relationship
\ben\label{1/NAnom}
\langle   \partial^{\mu} T_{\mu\nu}(0)  \psib(z_1)\psi(z_2) \rangle =
\frac{g^{4}}{4\pi}  \langle  \partial_{\nu}(\psib\psi)^{2}(0)
\psib(z_1)\psi(z_2) \rangle+ O(\sN) \ .
\een
Since the r.h.s. is not zero, it follows that the composite operator $T_{\mu\nu}$, as
defined, is {\em not} conserved. However, it follows that the operator
$\theta_{\mu\nu} := T_{\mu\nu} - (g^4/4\pi) \ \eta_{\mu\nu} (\psib \psi)^2$ {\em is} conserved
[up to order $O(\sN)$]. We consequently define $\theta_{\mu\nu}$ to be the {\em physical} stress energy tensor up to that order. Its VEV is found by taking the trace and using the now familiar OPE method, as
\ben\label{vev}
\langle \theta_{\mu\nu} \rangle = -\frac{1}{4 \pi \ell^2} \ \e^{-2\pi/g^2} \ \eta_{\mu\nu} + O(\N) \ .
\een
This corresponds to a negative vacuum energy of $\rho_\vac = -1/(4 \pi \ell^2) \ \e^{-2\pi/g^2}$ to leading order in $\N$. The negative sign is related to the negative 
sign of the $\beta$-function in the Gross-Neveu model.

We must finally discuss the {\em ambiguity} of our result. According to the general discussion
above, eq.~\eqref{amb}, we are still free to change $\theta_{\mu\nu} \to \theta_{\mu\nu} + \ell^{-2} c(g) \ \eta_{\mu\nu} {\bf 1}$, where $c(g) = c_0 + c_1 g + c_2 g^2 + \dots$ is {\em analytic}. This will result in
a corresponding change $\rho_\vac \to \rho_\vac + \ell^{-2} c(g)$. We can eliminate this remaining ambiguity by making the, reasonable sounding, assumption, that $\rho_\vac$ should vanish to all orders in perturbation theory. This is the same as demanding that, at the perturbative level, Minkowski space is
a solution to the semi-classical Einstein equations. Under this assumption $\rho_\vac= -1/(4 \pi \ell^2) \ \e^{-2\pi/g^2}$ is {\em unique}. This is the main result of this section.

\section{Conclusions}

We have defined the stress tensor as a composite operator which obeys the conservation law. Its expectation value in the vacuum state was found to be equal to~\eqref{vev}, corresponding to
the vacuum energy $\rho_\vac= -1/(4 \pi \ell^2) \ \e^{-2\pi/g^2}$.
The present model contains, as part of its definition at the quantum level, the dimensionful constant $\ell$ which corresponds to the units of length, and which are not provided by the classical Lagrangian~\eqref{Lagrange}. It would be more satisfactory to have a model wherein all dimensionful parameters are already part of the fundamental Lagrangian defining the theory in the ultra-violet. This can be achieved, in principle, by coupling our model to other fields with dimensionful couplings. For example, we could add
to the Lagrangian~\eqref{Lagrange} an interaction with some massive (by hand) scalar field $\varphi$ such as in
$
\mathcal{L} \to \mathcal{L} +\half \, \partial_\mu \varphi \partial^\mu \varphi + \half M^2 \,  \varphi^2 + y M  \, \varphi \overline \psi \psi,
$
where $M,y$ are new coupling parameters.
The constant $\ell$ can then be related to the dimensionful parameter $M$ by a renormalization condition, e.g. by demanding that
the physical (renormalized) mass of $\psi$ (as determined by the exponential decay of the $\psi$ 2-point function)
at some value $g = O(1) = y$ equals $M$, where $M$ is the physical (renormalized) mass of $\varphi$ (as determined by the exponential
decay of the $\varphi$ 2-point function). In a large $N$ analysis, one would also take $M=O(N)$. Although we will not carry out
such an analysis here, one would expect that the result for the vacuum energy is now modified to
$\rho_\vac \sim M^2 \e^{-O(1)/g^2}$, i.e. $\ell$ is simply set by $M$, which is now a parameter appearing explicitly in the Lagrangian. The renormalization ambiguity of $\theta_{\mu\nu}$ now consists in adding $c(g,y) M^2 \ \eta_{\mu\nu}$, where $c(y,g)$ is analytic. Again, we can eliminate this ambiguity by demanding that Minkowski space is a solution to the semi-classical Einstein equations to all perturbation orders in $y,g$, i.e. that $\langle \theta_{\mu\nu} \rangle$ vanishes to all orders in perturbation theory in $y,g$.
The real world, of course, is not described by the Gross-Neveu model, but by the Standard Model of elementary particle physics. However, if we pursue this analogy, $M$ would perhaps be replaced by a mass scale associated with the Standard Model Lagrangian, such as the Higgs mass $M \to M_{\rm H}$. Furthermore, the coupling would perhaps be replaced by a gauge coupling such as $g^2/4\pi  \to \alpha_{\rm EW} \sim \frac{1}{137}$. Assuming that these speculations are correct, we obtain an analog formula $\rho_\vac \sim M^4_{\rm H} \e^{-O(1)/\alpha_{\rm EW}}$ for some constant of order unity. The smallness of $\rho_\vac$ is
achieved by the characteristic non-perturbative dependence on the dimensionless coupling constant. In our model example, the sign of $\rho_\vac$ is negative, whereas vacuum energy
in our universe is positive. The sign in our model can be traced back
to the negative sign of the corresponding $\beta$-function. We do not know what the sign of $\rho_\vac$ may be in the Standard Model, but we note that there are
gauge couplings with either sign of the $\beta$-function.

To summarize, we believe that
non-perturbative effects are a potential explanation for the order of magnitude of Dark Energy.

\begin{acknowledgments}
This project was supported by Leverhulme Trust Grant no.~F/00407/BM. This work was begun during a stay of one of us (S.H.) at the Institute of Particle and Nuclear Studies, KEK, Japan in January 2010. S.H. gratefully acknowledges  hospitality and financial support. We would also
like to thank C. Kopper for explanations concerning the Gross-Neveu model.
\end{acknowledgments}
\bibliography{ccbib.bib}

\end{document}